\documentclass[preprint2,numberedappendix]{emulateapj}

\newcommand{\OII}{\ensuremath{{\rm [OII] \lambda 3727,3729} }}

\newcommand{\Msun}{\ensuremath{M_{\odot}}}
\newcommand{\Mdyn}{\ensuremath{M_\mathrm{dyn}}}

\newcommand{\Mstar}{\ensuremath{M_{\ast}}}
\newcommand{\HST}{\emph{HST}}
\newcommand{\spitzer}{\emph{Spitzer}}
\newcommand{\microm}{\ensuremath{{\rm \mu m } }}
\newcommand{\R}{\ensuremath{R_e}}
\newcommand{\Sersic}{S{\'e}rsic}

\newcommand{\sigmae}{\ensuremath{\sigma_e}}

\usepackage{capt-of}
\usepackage{graphicx}
\usepackage{subfigure}
\usepackage{verbatim}

\shorttitle{MOSFIRE Spectroscopy of $z>2$ Quiescent Galaxies}
\shortauthors{Belli et al.}
\submitted{Accepted for publication in ApJ Letters}

\begin{document}

\title{MOSFIRE Absorption Line Spectroscopy of $z>2$ Quiescent Galaxies: \\ Probing a Period of Rapid Size Growth}

\author{Sirio Belli\altaffilmark{1}, Andrew B. Newman\altaffilmark{2}, Richard S. Ellis\altaffilmark{1}, Nick P. Konidaris\altaffilmark{1}}
\altaffiltext{1}{Department of Astronomy, California Institute of Technology, MS 249-17, Pasadena, CA 91125, USA}
\altaffiltext{2}{The Observatories of the Carnegie Institution for Science, 813 Santa Barbara St., Pasadena, CA 91101, USA}

% ***************************************************************************************************
%						    ABSTRACT
% ***************************************************************************************************

\begin{abstract}
Using the MOSFIRE near-infrared multi-slit spectrograph on the Keck 1 Telescope, we have secured high signal-to-noise ratio absorption line spectra for six massive galaxies with redshift $2 < z < 2.5$. Five of these galaxies lie on the red sequence and show signatures of passive stellar populations in their rest-frame optical spectra. By fitting broadened spectral templates we have determined stellar velocity dispersions and, with broad-band HST and Spitzer photometry and imaging, stellar masses and effective radii. Using this enlarged sample of galaxies we confirm earlier suggestions that quiescent galaxies at $z>2$ have small sizes and large velocity dispersions compared to local galaxies of similar stellar mass. The dynamical masses are in very good agreement with stellar masses ($\log \Mstar/\Mdyn = -0.02 \pm 0.03$), although the average stellar-to-dynamical mass ratio is larger than that found at lower redshift ($-0.23 \pm 0.05$). By assuming evolution at fixed velocity dispersion, not only do we confirm a surprisingly rapid rate of size growth but we also consider the necessary evolutionary track on the mass-size plane and find a slope $\alpha = \mathrm{d} \log \R / \mathrm{d} \log \Mstar \gtrsim 2$ inconsistent with most numerical simulations of minor mergers. Both results suggest an additional mechanism may be required to explain the size growth of early galaxies.
\end{abstract}


% ***************************************************************************************************
%						1. INTRODUCTION
% ***************************************************************************************************

\section{Introduction}
\label{sec:intro}

The assembly history of nearby quiescent and morphologically early-type galaxies remains an important issue in extragalactic astronomy. Of particular interest is the fate of the population of compact red galaxies at redshift $z \sim 2$ \citep{daddi05, trujillo06, vandokkum06} which has been the subject of much observational effort. To match the properties of local galaxies, the growth in size must be significantly larger than the growth in mass. 

Although initial progress relied on photometric data, providing measures of both compact sizes and stellar masses of large samples beyond redshift $z \sim 1$ \citep[e.g.,][]{damjanov09, bezanson11, bezanson12}, key advances have become possible with spectroscopic samples. Spectroscopic data address the relative growth of the dynamical and stellar masses \citep{belli14lris}, as well as mean luminosity-weighted ages \citep{newman14}.  As stellar velocity dispersions should remain stable through merger episodes, spectroscopic observations can link high-redshift progenitors with their local descendants. This is particularly important in considering progenitor bias, i.e., the continued arrival of recently-quenched larger galaxies \citep[e.g.,][]{carollo13}. Via the first comprehensive spectroscopic sample at $z>1$, we quantified the size and mass growth rates of individual galaxies demonstrating significant growth \citep{belli14lris} at a rate consistent with minor mergers observed from independent imaging studies \citep{newman12}.

Attention now focuses on understanding the population of massive compact sources at $z > 2$. Deep imaging with the Wide Field Camera 3 (WFC3/IR) onboard \emph{Hubble Space Telescope} (\HST) has determined the growth rate is particularly rapid over the brief interval corresponding to $1.5<z<2.5$ \citep{newman12}. However, only limited spectroscopy is available as absorption line work is difficult in the near-infrared \citep[e.g.,][]{kriek06}. Until recently the relevant instruments (e.g. X-Shooter on the Very Large Telescope) were single-object long slit facilities. Despite heroic efforts, few stellar velocity dispersions are available beyond $z\simeq2$ \citep{vandokkum09, toft12, vandesande13}. 

The MOSFIRE multi-slit near-infrared spectrograph on Keck 1 \citep{mclean12} provides the first opportunity to systematically explore quiescent galaxies beyond $z \sim 2$. Here we present absorption line spectroscopy for a reasonable sample of compact massive galaxies at $2 < z < 2.5$. Our goal is to derive stellar velocity dispersions and dynamical masses, testing the rapid size growth rate inferred photometrically, as well as to examine this growth in the context of numerical simulations of galaxy merging.

\begin{figure*}[tbp]
\centering
\includegraphics[width=\textwidth]{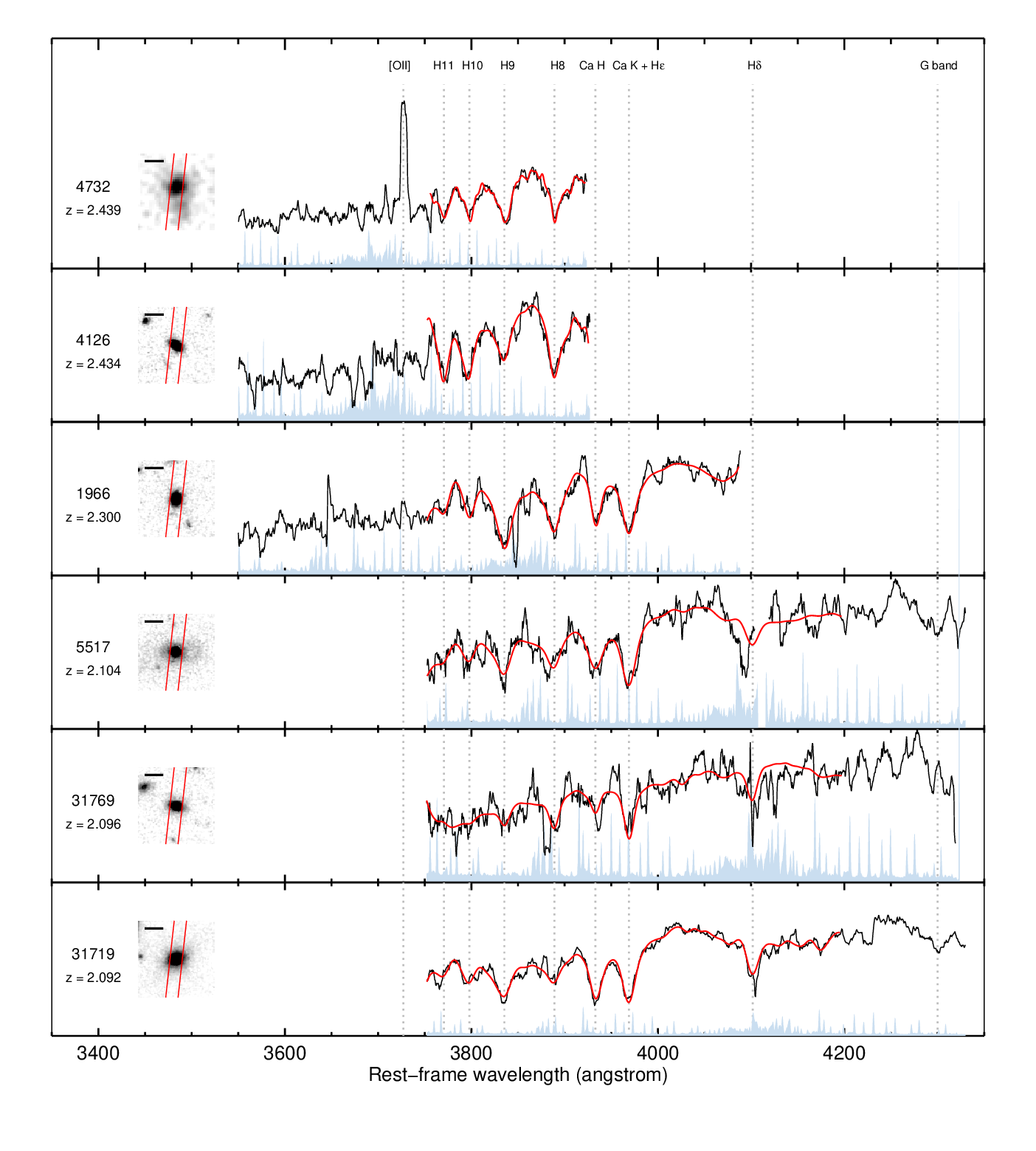}
\caption{\HST\ images and MOSFIRE spectra for a sample of six galaxies with detected absorption lines. For each object, the ID and spectroscopic redshift are indicated; the 4 arcsec cutout shows the F160W image with a 10 kpc ruler; the observed spectrum (inverse-variance smoothed, black line), its uncertainty (divided by 3, light blue), and the best-fit model (red line) are plotted. Absorption and emission features are marked by gray dotted lines. For 4732, only ground-based data are available, and the cutout is from the UltraVISTA H-band imaging \citep{mccracken12}.}
\label{fig:spectra}
\end{figure*}

Throughout  we use AB magnitudes and assume a $\Lambda$CDM cosmology with $\Omega_M$=0.3, $\Omega_{\Lambda}$=0.7 and $H_0$= 70 km s$^{-1}$ Mpc$^{-1}$.

% ***************************************************************************************************
%						2. DATA
% ***************************************************************************************************

\section{Data}
\label{sec:data}

\subsection{Target Selection and Ancillary Data}
\label{subsec:targetselection}

To select spectroscopic targets, we use public photometric data from the NEWFIRM Medium-Band Survey \citep[NMBS,][]{whitaker11}; this includes deep ground-based narrow, medium and broad band observations from the near-UV to the near-infrared and \spitzer\ IRAC and MIPS data. We limited our search to fields with high-quality \HST\ F160W imaging data from the Cosmic Assembly Near-IR Deep Extragalactic Legacy Survey \citep[CANDELS,][]{grogin11, koekemoer11}. 

We compiled a catalog of stellar masses, colors, and photometric redshifts derived via spectral energy distribution (SED) fitting. We selected the MOSFIRE pointing with the highest number of massive galaxies within the photometric redshift range $2 < z_\mathrm{phot} < 2.5$, using $U-V$ and $V-J$ rest-frame colors to prioritize quiescent objects \citep{williams09}. Our best pointing in the COSMOS field includes 21 targets, some of which belong to the $z \sim 2.1$ protocluster discovered by \citet{spitler12}.

\begin{deluxetable*}{lccccccccccccc}
\tabletypesize{\footnotesize}
\tablewidth{0pc}
\tablecaption{Properties of the MOSFIRE Sample \label{tab:sample}}
\tablehead{
\colhead{ID} & \colhead{R.A.} & \colhead{Decl.} & $J$ & $U-V$ & $V-J$ & \colhead{$z$} & \colhead{\sigmae} & \colhead{\R} & \colhead{$n$} & \colhead{$q$} & \colhead{$\log \Mstar/\Msun$} & \colhead{$\log \Mdyn/\Msun$}
\\
 & (J2000) & (J2000) & & & & & (km s$^{-1}$) & (kpc) & & & & 
} 
\startdata
31719 & 150.07393 & 2.2980 & 21.5 & 1.90 & 1.03 & 2.092 & $ 359 \pm 30 $ & 3.08 & 5.5 & 0.79 & $ 11.62 \pm 0.06 $ & $ 11.66 \pm 0.09 $  \\
31769 & 150.07460 & 2.3020 & 22.8 & 1.93 & 1.32 & 2.096 & $ 312 \pm 65 $ & 1.98 & 5.4 & 0.54 & $ 11.28 \pm 0.06 $ & $ 11.35 \pm 0.19 $  \\
5517 & 150.06562 & 2.2611 & 21.9 & 1.71 & 1.03 & 2.104 & \phantom{2}$ 464 \pm 138 $ & \phantom{\tablenotemark{a}}5.07\tablenotemark{a} & \nodata & \nodata & $ 11.41 \pm 0.06 $ & $ 12.10 \pm 0.17 $  \\
1966 & 150.05489 & 2.1982 & 22.5 & 1.75 & 0.96 & 2.300 & $ 350 \pm 61 $ & 1.01 & 2.7 & 0.71 & $ 11.22 \pm 0.06 $ & $ 11.16 \pm 0.16 $  \\
4126 & 150.05579 & 2.2361 & 22.8 & 1.57 & 0.70 & 2.434 & $ 223 \pm 56 $ & 1.25 & 1.4 & 0.45 & $ 10.89 \pm 0.04 $ & $ 10.86 \pm 0.22 $  \\
4732 & 150.05246 & 2.2455 & 22.1 & 1.36 & 1.19 & 2.439 & $ 128 \pm 48 $ & \nodata & \nodata & \nodata & $ 11.40 \pm 0.04 $ & \nodata
\enddata
\tablecomments{$U-V$ and $V-J$ are rest-frame colors. \sigmae\ is the velocity dispersion within one effective radius. The effective radius \R, \Sersic\ index $n$ and axis ratio $q$ are measured in F160W.}
\tablenotetext{a}{Size determined via curve of growth.}
\end{deluxetable*}

\subsection{Spectroscopic Data}
\label{subsec:specdata}

Spectroscopic observations were undertaken using MOSFIRE on April 17, 2013, and March 6, 7,  2014 with clear sky and 0.5-0.7 arcsec seeing. We observed in the J band using a two-point dithering pattern and an exposure time of 120 s per frame, for a total of 8hr20min. The 0.8 arcsec slit width yielded a spectral resolution of 45 km/s. The data were reduced using the MOSFIRE pipeline that performs flat fielding, sky subtraction and wavelength calibration, and outputs rectified 2-D spectra, from which 1-D spectra were optimally extracted. Telluric correction and flux calibration were performed using the spectra of A0 standard stars.

Continuum emission is detected in many spectra and six, shown in Figure \ref{fig:spectra}, exhibit absorption lines such as Ca II H and K and Balmer lines with well-defined redshifts in the range $2.092<z<2.439$ consistent with our photometric selection. Other features include the G band and the \OII\ emission line. The continuum signal-to-noise ratios range from 5 to 18 per resolution element. Table \ref{tab:sample} summarizes the observational data.

\subsection{The Target Sample}
\label{subsec:stellarpops}

\begin{figure}[bp]
\centering
\includegraphics[width=0.45\textwidth]{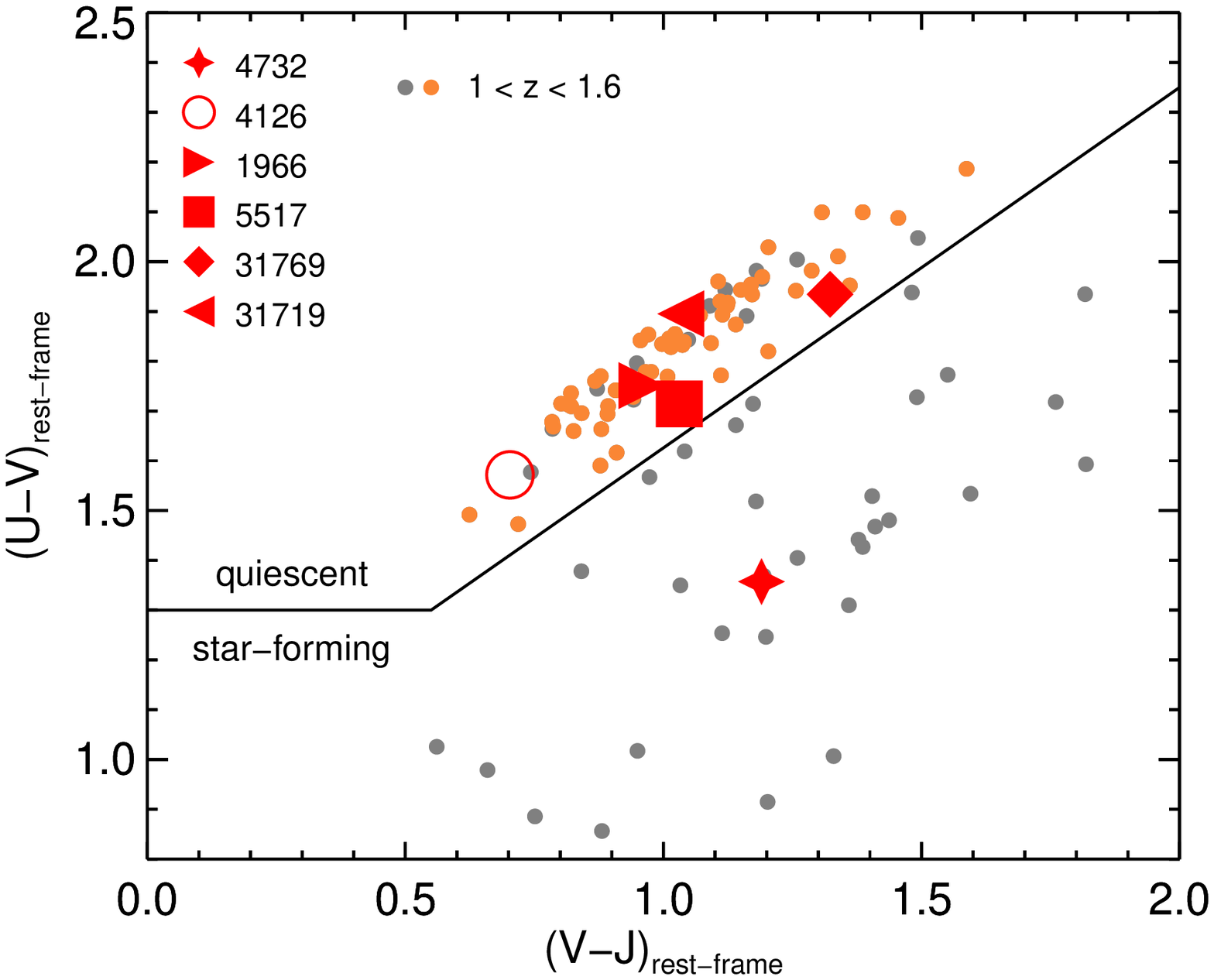}
\caption{The UVJ diagram. Large symbols represent our MOSFIRE sample. Small circles are lower-redshift ($1<z<1.6$) galaxies of \citet{belli14lris}, with orange indicating the quiescent subsample with velocity dispersion measurement (see Figures \ref{fig:cube} and \ref{fig:fixedsigma}).}
\label{fig:uvj}
\end{figure}

To gain insight into the nature of our sample we use the $UVJ$ diagram regularly employed to distinguish between star-forming and quiescent galaxies \citep[e.g.,][]{williams09} (Figure \ref{fig:uvj}).  Rest-frame $U-V$ and $V-J$ colors were calculated by integrating the best-fit SEDs (see Section \ref{subsec:fast}). We also show the $1<z<1.6$ sample of \citet{belli14lris} and the distinguishing line between quiescent and star-forming objects used in that work. Of the six selected galaxies, one (4732) is clearly star-forming given its [OII] emission, $UVJ$ colors and strong 24 \microm\ emission ($ 327 \pm 9$ $\mu$Jy). 4126 is likely recently-quenched: it lacks [OII] emission but shows strong Balmer absorption, much weaker 24 \microm\ emission ($28 \pm 8$ $\mu$Jy), and $UVJ$ colors of a post-starburst system. The other four galaxies show no signs of star formation: they have quiescent $UVJ$ colors, no [OII] emission, and are not detected at 24 \microm\ with the exception of 31769. For the latter, the mid-infrared emission ($161 \pm 11$ $\mu$Jy) is partly due to an active galactic nucleus (AGN), given its high X-ray luminosity \citep[$2.0 \pm 0.3 \cdot 10^{44} $ erg s$^{-1}$,][]{elvis09} and infrared colors \citep{stern05}. A more detailed study of the MOSFIRE spectra, including measurements of ages and star formation histories, will be presented in a future article (S. Belli et al., in prep.).

% ***************************************************************************************************
%						3. PHYSICAL PROPERTIES
% ***************************************************************************************************

\newpage

\section{Physical Properties}
\label{sec:properties}

\subsection{Structural Properties}
\label{subsec:galfit}

We derive structural parameters from the \HST\ F160W data that probe the rest-frame optical emission. We fit 2D \Sersic\ profiles using GALFIT \citep{peng02}, masking out neighboring objects and deriving the point spread function (PSF) from isolated bright stars. For each object the fitting procedure gives the \Sersic\ index $n$, the axis ratio $q$, and the circularized effective radius $\R = a \sqrt{q}$, where $a$ is the half-light semi-major axis (see Table \ref{tab:sample}). From the tests performed by \citet{newman12}, we estimate the uncertainty on the radii to be 10\%.

The \Sersic\ profile fit is good for all objects except 5517, which presents an asymmetric halo perhaps because of a merging event. This object lies at the center of the protocluster ``A'' described by \citet{spitler12}. We note that a luminous halo has been seen in at least another $z\sim2$ protocluster member \citep{strazzullo13}. For object 5517 alone we estimate the effective radius by performing a curve of growth analysis on the radial profile, after deconvolving the image with the PSF. We tested this method on the other targets and find agreement with the \Sersic\ fit to within 30\%. 

The object 4732 lies just outside the area covered by CANDELS, and no size measurement was attempted.
 
\begin{figure*}[htbp]
\vspace{10mm}
\centering
\includegraphics[width=\textwidth]{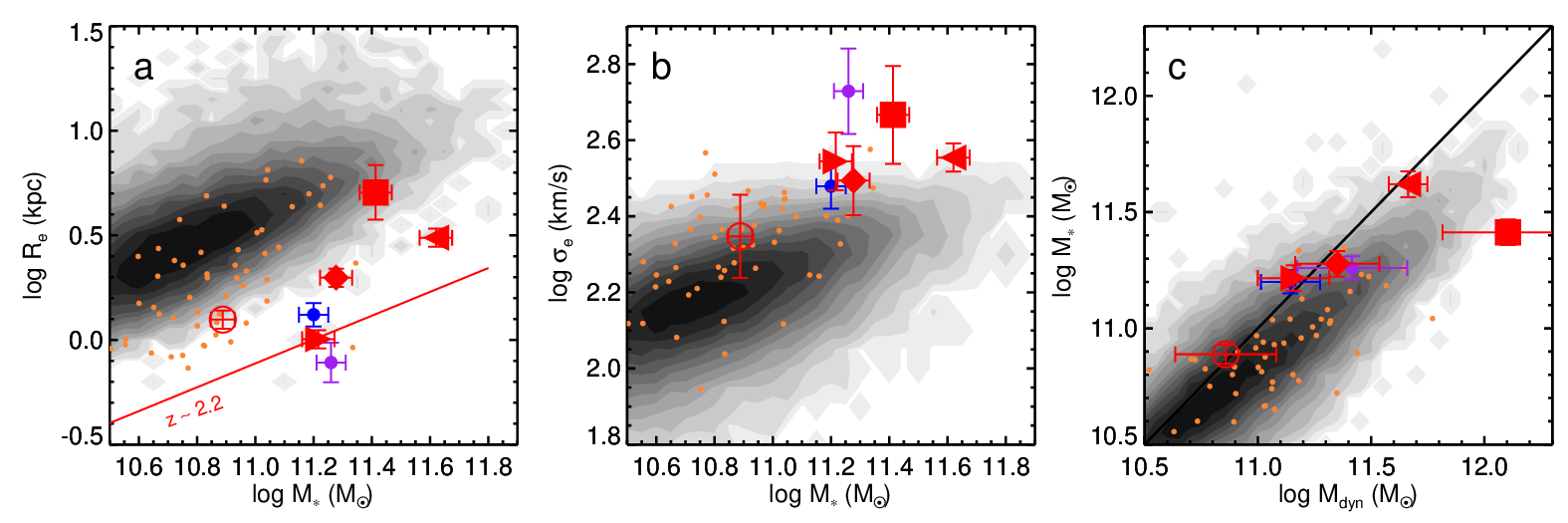}
\caption{Relations between stellar mass and structural and dynamical properties for quiescent galaxies with $0<z<2.5$. Samples include: a local population from SDSS (grayscale), the $1<z<1.6$ sample from \citet[][orange points]{belli14lris}, our MOSFIRE sample (red large symbols, as in Figure \ref{fig:uvj}), and a compilation of $z>2$ objects from the literature (\citealt[][purple]{vandokkum09}; \citealt[][blue]{vandesande13}). a) Mass-size relation. The red line indicates the $z \sim 2.2$ relation derived by \citet{newman12}. b) Mass-velocity dispersion relation. c) Dynamical versus stellar mass. The black line indicates the one-to-one relation.}
\label{fig:cube}
\end{figure*}

\subsection{Stellar Masses}
\label{subsec:fast}

We fit synthetic spectra to our extensive photometric data to measure stellar masses. We perform the fit using FAST \citep{kriek09} adopting the \citet{bruzual03} template library and the \citet{chabrier03} initial mass function (IMF). For the choice of the star formation history and stellar population parameters, we direct the reader to \citet{belli14lris}. To ensure consistency between the mass and size measurements, for each object we scale the observed SED to match the F160W flux obtained via \Sersic\ fit, following \citet{belli14lris}. The average correction is small, $-0.06 \pm 0.08$ dex.

The corrected stellar masses \Mstar\ and their uncertainties are listed in Table \ref{tab:sample}. The masses, which range from $10^{11}$ to $10^{11.6}$ \Msun, are very large, as expected from the combination of selection criteria and bias due to the absorption line detection.

\subsection{Velocity Dispersions}
\label{subsec:ppxf}

We measure the stellar velocity dispersions by fitting broadened templates to the MOSFIRE spectra in the range 3750-4200 \AA. We use the Penalized Pixel-Fitting (pPXF) code of \citet{cappellari04}, and take template spectra from the library of synthetic stellar populations by \citet{bruzual03}. Each spectrum is combined with an additive and a multiplicative polynomial to account for template mismatch and uncertainties in flux calibration or dust attenuation. Pixels contaminated by strong sky emission are excluded. The observed velocity dispersion is corrected for instrumental resolution, as measured from unblended sky lines, and template resolution.

We performed several tests to explore how robust are our derived velocity dispersions noting, in particular, the prevalence of Balmer absorption lines in some of the spectra. We followed the techniques discussed by van de Sande et al (2013) and Belli et al (2014) to which the reader is referred. We considered variations in the utilized wavelength range, fitting polynomial degree, template models and fraction of discarded pixels. We also studied the template dependence by comparing our results with those based on the higher resolution Indo-US library (Valdes et al 2004). For each object, the systematic error was determined by adding the contribution of these effects in quadrature to the random error. Finally, to achieve the dispersions within the effective radius \sigmae\ reported in Table \ref{tab:sample}, we apply a 5\% aperture correction \citep[based on the model by][]{vandesande13}. 

Object 31719 has already been observed with X-Shooter by \citet{vandesande13}, that derived $\sigmae = 446 ^{+54} _{-59}$ km s$^{-1}$. Our measurement is smaller, with a discrepancy at the 1.4$\sigma$ level. In the following, we will use our measurement for this object, since our spectrum has a higher signal-to-noise ratio.

% ***************************************************************************************************
%						4. DISCUSSION
% ***************************************************************************************************

\begin{figure*}[tbp]
\centering
\includegraphics[width=0.45\textwidth]{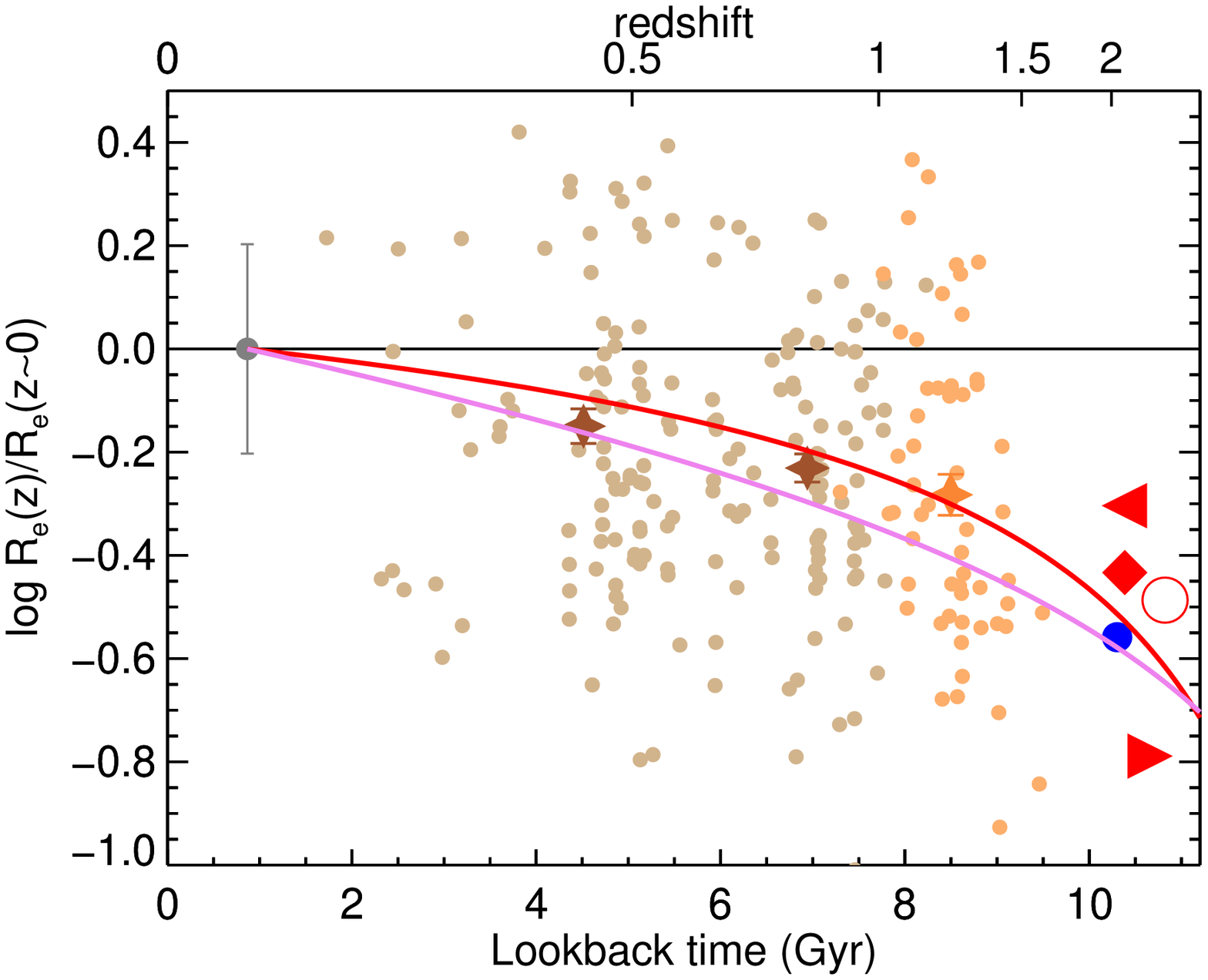}
\includegraphics[width=0.45\textwidth]{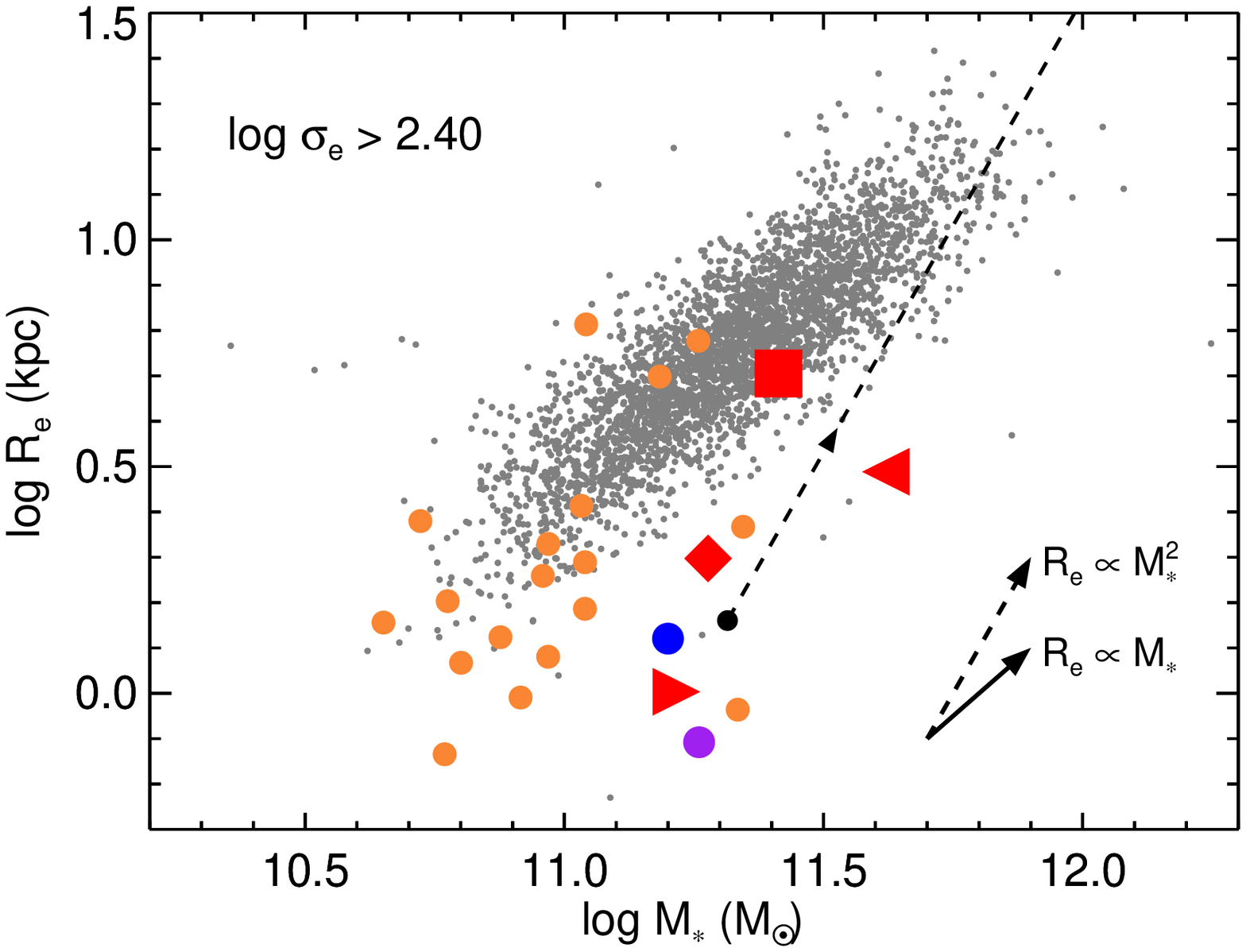}
\caption{Left: redshift evolution of effective radius at fixed velocity dispersion. The small points are a compilation of lower-redshift studies (\citealt[][brown]{treu05long}, \citealt[][orange]{belli14lris}), with the diamonds representing the running average. The red line is the fit to the evolution of the mass-size relation from \citet{newman12}, while the violet line is a fit to the observed evolution at fixed number density from \citet{vandokkum10}. Right: Mass-size plane for $0 < z < 2$ quiescent galaxies with $\log \sigmae > 2.40$. The high-redshift sample is clearly offset from the local population (gray points) toward small radii, at fixed stellar mass. Symbols as in Figures \ref{fig:uvj} and \ref{fig:cube}. The black point represents the mean of the high-redshift sample, and an evolutionary track with slope $\alpha=2$ is shown as a dashed line. The solid and dashed arrows represent the cases for $\alpha=1$ and $2$ respectively.}
\label{fig:fixedsigma}
\end{figure*}

\section{The Dynamical and Structural Evolution of Quiescent Galaxies}
\label{sec:discussion}

We use our new dynamical measurements, together with those from lower-redshift observations, to constrain the evolution of the size and structure of quiescent galaxies.

Figure \ref{fig:cube}a shows the mass-size relation for similarly-selected $UVJ$-quiescent galaxies over $0 < z < 2.5$. We show the local population from the Sloan Digital Sky Survey \citep[SDSS DR7,][grayscale map]{abazajian09}, a sample at $1 < z < 1.6$ from \citet[][small orange points]{belli14lris}, and the other two quiescent galaxies at $z>2$ for which velocity dispersion measurements have been published (\citealt[][purple]{vandokkum09}, and \citealt[][blue]{vandesande13}; see also \citealt{toft12}). Our MOSFIRE sample is shown in red. Clearly, the mass-size relation evolves with redshift. At fixed stellar mass, galaxies at $z \sim 1.3$ are about 0.25 dex smaller than the local population. At $z>2$, the logarithmic offset from the local sample nearly doubles, implying evolution accelerates at earlier cosmic times. These findings confirm the results of previous photometric studies, as shown by the agreement of our points with the mass-size relation at $z\sim 2.2$ from \citet[][red line]{newman12}. Object 5517, the brightest galaxy (BCG) of a protocluster, is an exception, suggesting that such systems have large sizes already at $z\sim2$, in agreement with other studies at $z\lesssim1.8$ \citep{papovich12, stanford12, newman14}.

The main advance of this paper is that we can now explore the dynamical properties of galaxies at $z > 2$. Figure \ref{fig:cube}b shows high-redshift galaxies have significantly larger velocity dispersions than lower redshift objects of similar stellar mass. Most of our MOSFIRE objects have similar dispersions, $\sigmae \approx 300$ km s $^{-1}$, with the exception of 4126 (shown as an open symbol). This galaxy is the only post-starburst object in our sample, and presents an elongated morphology and low \Sersic\ index, $n=1.4$, typical of disk-like galaxies.

Velocity dispersions enable us to calculate dynamical masses, via $\Mdyn = 5 \sigmae^2 \R / G$. Figure \ref{fig:cube}c compares the dynamical and stellar masses, again contrasting the trend with samples drawn from the literature. While the $z\sim1.3$ sample closely follows the local distribution, $z>2$ galaxies tend to have higher stellar-to-dynamical mass ratios. This difference was first suggested by \citet{toft12} and \citet{vandesande13}, but as our MOSFIRE sample doubles the number of dynamical masses with $z>2$, it is now more significant, particularly as our velocity dispersions are more accurate.

We calculate the average mass ratio for all the $z>2$ quiescent galaxies, excluding the BCG, and we find $\log \Mstar/\Mdyn = -0.02 \pm 0.03$ and an average stellar mass of $10^{11.3} \Msun$. Although our sample is modest, this is a remarkably tight agreement. Considering the most massive galaxies of the $1<z<1.6$ sample, we find $\log \Mstar/\Mdyn = -0.23 \pm 0.05$ and an average mass of $10^{11.1} \Msun$. This significant evolution could arise if $z>2$ quiescent galaxies have a reduced dark matter fraction, a heavier stellar IMF or different structure compared to their lower-redshift counterparts.

We now use our dynamical masses to infer the rate of size growth for high-redshift quiescent galaxies. Following \citet{belli14lris}, we assume that we can link progenitor and descendant galaxies by selecting populations at \emph{fixed velocity dispersion}. This follows the results of numerical simulations that show that velocity dispersion is minimally affected during merger events \citep[e.g.][]{hopkins09scalingrel,oser12}, and the observed unchanging velocity dispersion function \citep{bezanson12}. For each of our $z>2$ objects, we therefore select SDSS galaxies with similar velocity dispersions (within 0.05 dex). We calculate the median size of the local subsample and assume that our high redshift object will grow in size until it reaches that value. We repeat this procedure for lower-redshifts samples and show the inferred growth rates in the left panel of Figure \ref{fig:fixedsigma}. Typical $z>2$ galaxies\footnote{The BCG and the \citet{vandokkum09} galaxy are omitted because in the local universe there are no objects with such high velocity dispersions. Additionally both have the most uncertain dispersions.} are noticeably smaller than those at $z\sim1.3$, despite the small interval in cosmic time between the two samples. Our results are qualitatively in agreement with the size growth at \emph{fixed number density} derived by \citet[][violet line]{vandokkum10}. Interestingly, the growth of individual galaxies is not dissimilar to that of the total population derived from the evolution of the mass-size relation \citep[][red line]{newman12}. Although minor merging is likely responsible for the size evolution of quiescent galaxies at $z<1.5$, such a rapid growth at $z\sim2$ is hard to reconcile with the observed merger rate (\citealt{newman12}; see also \citealt{cimatti12}).

In the right panel of Figure \ref{fig:fixedsigma} we show the mass-size relation for galaxies with $\log \sigmae > 2.40$ at all redshifts. This velocity dispersion bin includes all the $z>2$ galaxies except the post-starburst object 4126. The local population forms a clear sequence whereas objects at intermediate redshifts show a mild offset towards smaller sizes (and masses). However, the $z>2$ galaxies populate a distinct region of the mass-size plane with almost no overlap with lower redshift samples; almost all have radii 0.3 dex below the local sample. The inescapable conclusion is that quiescent galaxies at $z>2$ must undergo a dramatic and rapid size growth.

A powerful method to constrain the physical processes responsible for this size growth is to measure the slope $\alpha = \mathrm{d} \log \R / \mathrm{d} \log \Mstar $ of the evolutionary tracks on the mass-size plane and compare it with theoretical predictions. Simple virial arguments \citep{naab09, bezanson09} give $\alpha = 1$ for identical mergers and $\alpha = 2$ for the limiting case of mergers with infinitely diffuse satellites. More realistic numerical simulations, which include the effect of dark matter, gas, and a distribution of orbits, indicate that minor mergers are less efficient than the theoretical limit, and yield values in the range $1.4 < \alpha < 1.8$ \citep{hopkins09scalingrel, nipoti12, oser12, posti14}. The simulations of \citet{hilz13}, in which massive dark matter halos enhance the efficiency of minor merging up to $\alpha=2.4$, are the only exception. However, the large dark matter fraction at the center of these simulated galaxies disagrees with the observed stellar-to-dynamical mass ratios at both low and high redshift.

Assuming evolution at fixed velocity dispersion, we measure $\alpha$ by considering the tracks that high-redshift points must follow in order to match the local distribution. Using this technique, the $z\sim1.3$ sample yields $\alpha = 1.4 \pm 0.2$ \citep{belli14lris}. Merging can therefore readily explain the size growth over $0 < z < 1.5$, a conclusion supported by direct imaging \citep{newman12}. However, at $z>2$ the growth is clearly much more rapid. It is not possible to derive a robust measurement of the slope $\alpha$ for two reasons: firstly, our $z>2$ sample is not velocity dispersion complete; secondly, one of our basic assumptions might not hold, since the velocity dispersion function has not been probed beyond $z\sim1.5$. As the number density of quiescent galaxies declines steeply at this redshift, a strong progenitor bias is expected. Despite these limitations, we can still derive an important lower limit on $\alpha$, by assuming that the average high-redshift galaxy (shown in black, excluding the BCG) will evolve into one of the most massive objects at $z\sim0$. Using this method, for all the $z>2$ quiescent galaxies excluding the BCG we derive a lower limit of $\alpha \gtrsim 2$, shown as a dashed line in the figure. 

In summary, our spectroscopic data allows us to conclude that both the absolute rate of size growth {\it and} the inferred motion in the mass-size plane are independently inconsistent with minor mergers being the principal physical process governing the evolution of quiescent galaxies at $z\sim2$. 
\\

We thank Chuck Steidel, Ian McLean and their team for their work in producing the remarkable MOSFIRE instrument. We thank Guillermo Barro for useful discussions. The authors acknowledge the very significant cultural role that the summit of Mauna Kea has always had within the indigenous Hawaiian community. We are most fortunate to have the opportunity to conduct observations from this mountain.

% ***************************************************************************************************
%							BIBLIOGRAPHY
% ***************************************************************************************************

\end{document}